\documentclass[cameraready]{Interspeech}
\usepackage{booktabs}    
\usepackage{multirow}    
\usepackage{makecell}    
\usepackage{graphicx}    
\usepackage{amsmath}     
\usepackage{adjustbox}
\usepackage{caption}
\usepackage{float}
\usepackage{stfloats}
\usepackage{siunitx}
\usepackage{lineno}

\title{Discrete vs. Continuous: A Comprehensive Study of Unified Audio Understanding in LALMs}

\author[affiliation={1}, orcid=0009-0000-5403-8432, equalcontribution]{Jing}{Peng}
\author[affiliation={1}, orcid=0009-0000-6247-6379, equalcontribution]{Zichao}{Nie}
\author[affiliation={1},orcid=0009-0002-0066-8214]{Zhisheng}{Zhang}
\author[affiliation={1},orcid=0009-0007-2050-263X]{Jingran}{Xie}
\author[affiliation={1}, orcid=0000-0001-8533-0524, correspondingauthor]{Zhiyong}{Wu}

\address{
    $^1$ Shenzhen International Graduate School, Tsinghua University, Shenzhen, China 
}

\email{zywu@se.cuhk.edu.hk}

\keywords{large audio language models, audio encoders, benchmarking, unified audio understanding}

\usepackage{comment}

\begin{document}

\maketitle

\begin{abstract}
Large Audio Language Models (LALMs) utilize either continuous features or discrete tokens, yet the optimal representation paradigm for general audio understanding remains debated. Existing benchmarks often focus on narrow domains or evaluate encoders outside LALM contexts. To address these gaps, we systematically evaluate continuous and discrete representations across speech, sound and music. Utilizing our UniARC framework with dual evaluation strategies across model scales from SmolLM2-135M to Llama-3-8B, we analyze the dynamic relationships of data volume, model capacity, and computational efficiency. Our results reveal the pivotal role of semantic constraints in tokenization for audio understanding and demonstrate that scaling backbones fail to compensate for information loss in audio representation, especially in data-limited tasks. These findings offer practical guidance for balancing semantic density, fidelity, and efficiency in future LALMs.
\end{abstract}

\section{Introduction}

In recent years, Large Audio Language Models (LALMs) have advanced significantly~\cite{chu2023qwen,tang2023salmonn}. The underlying representation schemes of these models have undergone several developments, evolving from continuous features to discrete tokens, and currently towards hybrid approaches where both paradigms coexist or are integrated~\cite{wang2025speech,xu2024comparing,zhou2025voxcpm}. Discrete representations, owing to their structural consistency with text tokens, have greatly simplified the process of cross-modal alignment~\cite{borsos2023audiolm,zhang2023speechgpt}. However, the quantization bottleneck inherent in discretization often leads to the loss of crucial acoustic details, prompting researchers to re-examine the value of continuous features.

While numerous studies have emerged in this field, they often exhibit specific limitations. Some works focus on comparing semantic discrete tokens with continuous features, neglecting acoustic-based discrete representations~\cite{wang2025speech, xu2024comparing}. Others conduct evaluations outside the context of large language models (LLMs)~\cite{mousavi2024dasb, zhang2025x}, or concentrate on discrete tokens without a comprehensive benchmark against continuous features~\cite{mousavi2025discrete, guo2025recent}. 
Notably, most research focuses narrowly on speech, overlooking general audio understanding~\cite{wang2025speech}. Furthermore, existing evaluations are often confined to constrained or ``toy'' settings, neglecting the complex interplay between representation paradigms, model capacity, and data scalability~\cite{xu2024comparing}. 
Therefore, a systematic comparison addressing both cross-domain capabilities and real-world scaling dynamics is urgently needed.

To bridge these gaps, we propose the \textbf{Uni}fied \textbf{A}udio \textbf{R}epresentation \textbf{C}omparison (\textbf{UniARC}) framework, specifically designed for audio understanding. 
In this study, we unify the terminology by referring to audio representation modules as encoders, encompassing continuous features from Self-Supervised Learning (SSL), and discrete tokens derived from either clustering-based or neural codec-based paradigms. To ensure the robustness of our findings and simulate real-world deployment constraints, UniARC evaluates encoders under two distinct strategies: (1) a parameter-efficient fine-tuning setup leveraging SmolLM2-135M/360M~\cite{allal2025smollm}, and (2) a frozen-backbone probing pipeline using Llama-3-1B/8B~\cite{grattafiori2024llama}. 
Beyond a direct comparison of encoding paradigms, we leverage the UniARC to investigate the multi-dimensional interplay between data volume, model capacity, and computational efficiency.

Our analysis reveals that regardless of the representational form, the efficacy of audio encoders for understanding tasks is primarily determined by the richness of their semantic information. Specifically, we find that discrete tokens can surpass continuous features when they incorporate strong semantic constraints, whereas high-fidelity representations focused purely on signal reconstruction often fail in understanding tasks. These findings suggest that the semantic compatibility of audio representations with the language model's reasoning space is the primary determinant of performance in LALMs. Furthermore, we demonstrate that for audio understanding, scaling the language backbone cannot effectively compensate for insufficient representational quality. Our findings reveal that the front-end features effectively set a performance ceiling that remains unchanged or even declines despite increases in model scale. The primary contributions of this work are as follows:
\begin{itemize}

\item We comprehensively evaluate across three diverse audio domains, i.e., speech, general sound, and music, to compare LLM-based continuous and discrete representations specifically for audio understanding tasks.

\item We analyze how representation paradigms, model scales and data scalability interact, demonstrating that for audio understanding, aligning audio encoders with textual semantics is more decisive than scaling LLM backbones.

\item We identify the task-dependent strengths of different paradigms within the understanding context, providing guidelines for balancing semantic density, acoustic fidelity, and training efficiency in future LALMs designs.

\end{itemize}

\section{Evaluation Methodology}

\subsection{Evaluation Framework}

\begin{figure*}[!t]  
  \centering
  \includegraphics[width=0.95\textwidth]{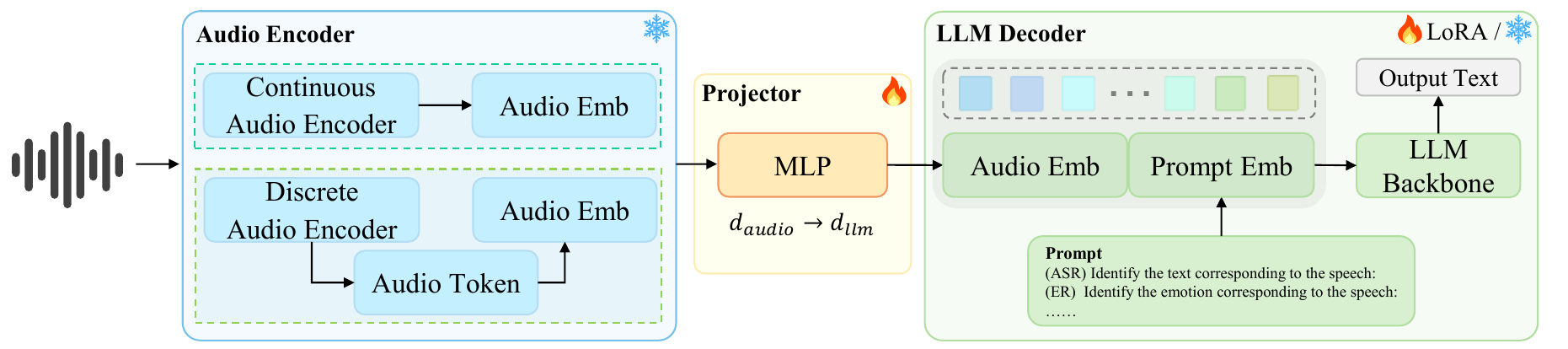} 
  \caption{The UniARC evaluation framework. The modular design supports flexible configurations of encoders, projectors, and LLM backbones. Fire and snowflake icons represent the training and frozen status of components, respectively.}
  \label{fig:speech_production}
\end{figure*}

Building upon the open-source XARES-LLM project~\cite{zhang2025x}, we propose \textbf{UniARC}, a modular framework that systematically evaluates audio encoders via an instruction-tuning paradigm, as shown in Figure~\ref{fig:speech_production}. Audio embeddings are appended to task-specific prompts, guiding the LLM within a unified sequence-to-sequence generation framework. To simulate diverse real-world deployment scenarios, we implement two strategies.

\textbf{The first scenario targets domain-specific experts via parameter-efficient fine-tuning.} For lightweight applications where backbone updates are computationally feasible, we employ SmolLM2 (135M/360M) with a LoRA-based instruction-tuning strategy~\cite{hu2022lora}, which updates only the adapters while mapping audio embeddings via a projector, maximizing performance on specific tasks within limited parameter budgets.

\textbf{The second scenario evaluates general-purpose LLMs via frozen-backbone probing.} To preserve inherent reasoning capabilities and minimize confounding variables from model adaptation, we utilize Llama-3 (1B/8B) with frozen parameters. To bridge the modality gap without updating the LLM, we employ a robust two-layer MLP projector and a 10-frame temporal concatenation to handle long acoustic sequences efficiently. This design isolates the adaptive capacity of LLM, directly exposing the intrinsic semantic quality of frontend encoders.

\subsection{Processing of Audio representations}




To ensure fair comparison, UniARC standardizes the conversion of diverse audio inputs into a unified format compatible with the LLM projector.

For \textbf{continuous representations}, raw audio is processed by encoders into high-dimensional dense embeddings, preserving representational density within a continuous latent space. For \textbf{discrete representations} derived from semantic clustering (K-means) or neural codecs, inputs are first converted to token indices and then mapped to codebook embeddings, effectively transforming discrete units into a continuous vector space.

In both paradigms, the resulting embeddings are processed by a projector to align dimensions with the LLM's text space, enabling end-to-end differentiable training. Finally, task-specific prompts guide the model across all downstream tasks.

\begin{figure*}[tbp]
    \centering
    \includegraphics[width=\linewidth]{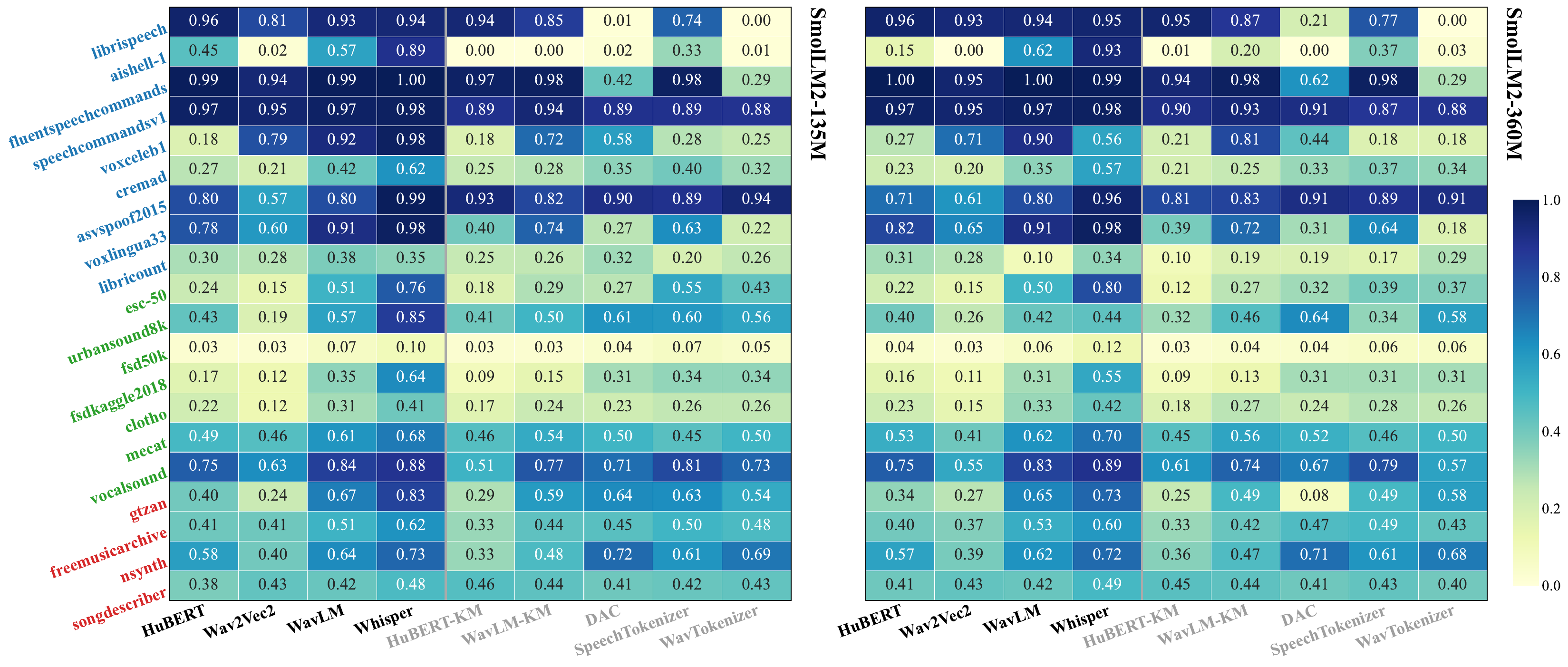}%
    \caption{Performance of representation paradigms within the UniARC fine-tuning pipeline (KM stands for K-means).}
    \label{fig:smollm2-heatmap}
\end{figure*}

\section{Experiments}

\subsection{Encoder Selection}
We include a diverse set of encoders to cover different representation paradigms. For continuous features, we select HuBERT~\cite{hsu2021hubert}, WavLM~\cite{chen2022wavlm}, Wav2Vec 2.0~\cite{baevski2020wav2vec}, and Whisper~\cite{radford2023robust} to represent SSL and weakly supervised approaches.

For discrete representations, we investigate two primary categories of encoders based on discretization mechanisms: 1) Clustering-based tokens, derived via K-means (1,000 centroids) on SSL features (HuBERT, WavLM) to prioritize high-level semantic abstraction~\cite{wang2025speech}; and 2) Neural Codec tokens, learned through end-to-end reconstruction. For the latter, we evaluate DAC~\cite{DAC} and WavTokenizer~\cite{ji2024wavtokenizer}, alongside SpeechTokenizer~\cite{zhang2023speechtokenizer}, a specialized codec that integrates semantic distillation to decouple semantic content from acoustic details.

To ensure consistent feature quality, we operate all encoders at 16~kHz, with the exception of WavTokenizer, which utilizes its native 24~kHz configuration. For continuous encoders (HuBERT, WavLM), we consistently extract the last hidden state.
For clustering-based discrete tokens, we set K-means clusters to 1000 for SmolLm2-135M/360M lightweight decoders with LoRA fine-tuning, balancing speech unit complexity, training stability, inference efficiency and computational cost~\cite{wang2025speech}.

\subsection{Tasks and Datasets}

To align with the two evaluation strategies defined in UniARC, we adopt a stratified task selection protocol covering Speech, Sound, and Music. A comprehensive overview of all evaluated datasets, task types, and metrics is summarized in Table \ref{tab:datasets}.

\begin{table}[th]
\caption{Overview of datasets, tasks, and metrics utilized in the UniARC framework. \textbf{Mode} denotes the evaluation setup: Tuned (\textbf{T}) with SmolLM2, Frozen (\textbf{F}) with Llama-3, or \textbf{Both}.}
\label{tab:datasets}
\centering
\resizebox{\columnwidth}{!}{
\begin{tabular}{@{}lllc@{}}
\toprule
\textbf{Dataset} & \textbf{Task Type (Abbr.)} & \textbf{Metric} & \textbf{Mode} \\
\midrule
\multicolumn{4}{@{}c@{}}{\textbf{\textit{Speech}}} \\
\midrule
LibriSpeech (LS)~\cite{panayotov2015librispeech} & Speech Recognition (ASR) & WER/iWER & Both \\
AISHELL-1~\cite{bu2017aishell} & Speech Recognition (ASR) & iCER & T \\
Fluent Speech Commands~\cite{ravanelli2019speech} & Intent Classification (IC) & Acc & T \\
SLURP~\cite{bastianelli2020slurp} & Intent Classification (IC) & Acc & F \\
Speech Commands V1~\cite{warden2018speech} & Keyword Spotting (KS) & Acc & T \\
CREMA-D~\cite{cao2014crema} & Emotion Recognition (ER) & Acc & Both \\
IEMOCAP~\cite{busso2008iemocap} & Emotion Recognition (ER) & Acc & F \\
VoxCeleb1~\cite{nagrani2019voxceleb} & Speaker Identification (SI) & Acc & T \\
VoxLingua33~\cite{Valk2020VOXLINGUA107AD} & Language Identification (LID) & Acc & T \\
LibriCount~\cite{stoter2018libricount} & Speaker Counting (SC) & Acc & T \\
VocalSound~\cite{gong2022vocalsound} & Human Sound Classification (HSC) & Acc & T \\
ASVspoof2015~\cite{wu2017asvspoof} & Spoofing Detection (SD) & Acc & T \\
\midrule
\multicolumn{4}{@{}c@{}}{\textbf{\textit{Sound}}} \\
\midrule
ESC-50~\cite{piczak2015esc} & Environment Classification (EnvC) & Acc & T \\
UrbanSound8K~\cite{salamon2014dataset} & Urban Sound Classification (USC) & Acc & Both \\
FSD50K~\cite{fonseca2021fsd50k} & Sound Event Detection (SED) & mAP & T \\
FSDKaggle2018~\cite{fonseca2018general} & Sound Event Detection (SED) & mAP & T \\
Clotho~\cite{drossos2020clotho} & Sound Captioning (SCap) & FENSE & Both \\
MECAT~\cite{niu2025mecat} & General Caption (GCap) & DATE & T \\
\midrule
\multicolumn{4}{@{}c@{}}{\textbf{\textit{Music}}} \\
\midrule
GTZAN~\cite{tzanetakis2001automatic} & Genre Classification (GC) & Acc & Both \\
Free Music Archive (FMA)~\cite{Defferrard2016FMAAD} & Genre Classification (GC) & Acc & T \\
NSynth~\cite{engel2017neural} & Instrument Classification (InstC) & Acc & T \\
Song Describer (SDD)~\cite{manco2023song} & Music Captioning (MCap) & FENSE & Both \\
\bottomrule
\end{tabular}
}
\end{table}

For parameter-efficient fine-tuning evaluations, we utilize a broad suite of instruction-tuning tasks adhering to XARES-LLM to establish a fundamental baseline. Conversely, for frozen-backbone probing evaluations, we strategically curate a representative subset of core tasks. Notably, SLURP and IEMOCAP are introduced here to rigorously probe encoders' intrinsic semantic reasoning.

We employ metrics tailored to each task type: Accuracy (Acc) for single-label classification, mAP for multi-label event detection, FENSE~\cite{zhou2022can} for audio/music captioning to evaluate semantic fluency, and DATE~\cite{niu2025mecat} for general captioning. For speech recognition, standard Word/Character Error Rates (WER/CER) are calculated. To facilitate intuitive visualization of fine-tuning evaluations, WER/CER are inverted into Inverse Word/Character Error Rate (iWER/iCER), ensuring that higher scores consistently indicate better performance.

\subsection{Experimental Setup}


Our experiments are conducted on NVIDIA A100 GPUs using PyTorch. For Fine-Tuning Strategy, we apply LoRA ($r=8, \alpha=32$) to LLM backbone on single GPU, maintaining consistent default hyperparameters on a single GPU. For Frozen-Probing Strategy, we freeze LLM and optimize only projector using AdamW optimizer~\cite{loshchilov2018decoupled} with a learning rate of $1 \times 10^{-5}$ across four GPUs. To ensure  fair comparison of representation quality, all task-specific configurations within each strategy are kept strictly identical across different audio encoders.

\section{Results and Analysis}

We comprehensively evaluate continuous features and discrete tokens across diverse tasks under UniARC framework. Results from the parameter-efficient fine-tuning strategy are visualized in Figure~\ref{fig:smollm2-heatmap}, where datasets are color-coded by domain(speech in blue, sound in green, music in red) on y-axis, and encoders are color-coded by representation type (continuous in black, discrete in gray). These results are numerically summarized in Table~\ref{tab:encoder_decoder_vertical} by averaging the scores within each domain. Conversely, detailed findings from the strict frozen-backbone probing strategy are presented in Table~\ref{tab:main_results}. 
The following subsections systematically analyze these results regarding representation paradigms, scaling effects, and training efficiency.

\subsection{Analysis of Representation Paradigms }

Based on the experimental results in Figure~\ref{fig:smollm2-heatmap} and Table~\ref{tab:main_results}, we find that whether an encoder is discrete or continuous is not the primary factor governing performance; rather, semantic compatibility fundamentally dictates its efficacy for LLM-based understanding, far outweighing raw signal fidelity.

\begin{table}[t]
    \centering
    \caption{Encoder Performance on Different Backbones. Higher scores indicate better performance ($\uparrow$).}
    \footnotesize  
    \setlength{\tabcolsep}{3pt}  
    \setlength{\heavyrulewidth}{0.8pt}  
    \setlength{\lightrulewidth}{0.4pt}  
    \begin{tabular}{l *{4}{S[table-format=1.3]}}  
      \toprule  
      \textbf{Encoder}               & \textbf{Speech} & \textbf{Sound} & \textbf{Music} & \textbf{Overall} \\
      \midrule  
      \multicolumn{5}{c}{\textit{SmolLM2-135M}} \\
      \midrule  
      HuBERT          & 0.634    & 0.333   & 0.445   & 0.491     \\
      Wav2Vec2         & 0.575    & 0.243   & 0.370   & 0.418     \\
      WavLM           & 0.766    & 0.466   & 0.561   & 0.620     \\
      Whisper               & \bfseries 0.858    & \bfseries 0.618   & \bfseries 0.666   & \bfseries 0.735     \\
      HuBERT+Kmeans         & 0.534    & 0.263   & 0.356   & 0.403     \\
      WavLM+Kmeans          & 0.620    & 0.361   & 0.488   & 0.503     \\
      DAC                   & 0.419    & 0.381   & 0.554   & 0.432     \\
      SpeechTokenizer       & 0.593    & 0.440   & 0.541   & 0.529     \\
      WavTokenizer          & 0.353    & 0.408   & 0.535   & 0.409     \\
      \midrule  
      \multicolumn{5}{c}{\textit{SmolLM2-360M}} \\
      \midrule  
      HuBERT          & 0.603    & 0.331   & 0.432   & 0.474     \\
      Wav2Vec2         & 0.589    & 0.238   & 0.368   & 0.422     \\
      WavLM           & 0.732    & 0.438   & 0.555   & 0.594     \\
      Whisper               & \bfseries 0.807    & \bfseries 0.561   & \bfseries 0.635   & \bfseries 0.687     \\
      HuBERT+Kmeans         & 0.502    & 0.304   & 0.366   & 0.407     \\
      WavLM+Kmeans          & 0.665    & 0.407   & 0.457   & 0.533     \\
      DAC                   & 0.435    & 0.391   & 0.417   & 0.416     \\
      SpeechTokenizer       & 0.583    & 0.375   & 0.505   & 0.494     \\
      WavTokenizer          & 0.347    & 0.378   & 0.523   & 0.393     \\
      \bottomrule  
    \end{tabular}
    \label{tab:encoder_decoder_vertical}
\end{table}

This principle is evident among continuous encoders. Weakly supervised Whisper aligns features with textual semantics, leads across most benchmarks. Conversely, Wav2Vec2's self-supervised focus on low-level acoustics limits its efficacy in non-speech semantic tasks. This confirms that while SSL models excel at signal modeling, Whisper's text-aligned features are more effective for high-level LLM reasoning.

A similar trend governs discrete representations, explaining the divergence among tokenization paradigms. While K-means quantization typically lags behind continuous features due to information loss, encoders that incorporate explicit semantic constraints defy this trend. Notably, SpeechTokenizer consistently surpasses its continuous teacher, HuBERT, in specific tasks such as ER. This demonstrates that semantically-aware discretization can act as a regularizer, distilling high-density semantic features optimally tailored for LLM comprehension.

Conversely, reconstruction-oriented codecs (e.g., DAC and WavTokenizer) struggle with semantic understanding tasks despite their superior fidelity. The preservation of low-level acoustic details, such as phase and microstructural information, may introduce irrelevant noise that complicates the mapping of audio features into the LLM's semantic space. However, this high fidelity remains valuable for timbre-sensitive tasks like music classification (NSynth), where fine-grained frequency details are more critical than high-level semantic abstractions.

\begin{table*}[!t]
  \centering
  \caption{Experimental results across Speech, Sound, and Music domains. Best results within each model scale (1B/8B) are \textbf{bolded}.}
  \label{tab:main_results}
  \resizebox{\textwidth}{!}{
  \begin{tabular}{l | ccccc | cc | cc}
    \toprule
    \multirow{3}{*}{\textbf{Model}} & \multicolumn{5}{c|}{\textbf{Speech}} & \multicolumn{2}{c|}{\textbf{Sound}} & \multicolumn{2}{c}{\textbf{Music}} \\
    \cmidrule(lr){2-6} \cmidrule(lr){7-8} \cmidrule(lr){9-10}
    
    & \multicolumn{2}{c}{ASR (WER \% $\downarrow$)} & \multicolumn{2}{c}{ER (ACC \% $\uparrow$)} & IC (ACC \% $\uparrow$) & USC (ACC \% $\uparrow$) & SCap (FENSE $\uparrow$) & GC (ACC \% $\uparrow$) & MCap (FENSE $\uparrow$) \\
    \cmidrule(lr){2-3} \cmidrule(lr){4-5} \cmidrule(lr){6-6} \cmidrule(lr){7-7} \cmidrule(lr){8-8} \cmidrule(lr){9-9} \cmidrule(lr){10-10}
    
    & \makecell{LS-960 \\ (clean/other)} & \makecell{LS-100 \\(clean)} & IEMOCAP & CREMA-D & SLURP & UrbanSound8K & Clotho & GTZAN & SDD \\
    \midrule
    
    \multicolumn{10}{c}{\textit{1B Models}} \\
    \midrule
    HuBERT     & \textbf{2.31/4.62} & \textbf{3.45} & 52.30 & 50.12 & \textbf{71.23} & 55.56 & 0.196 & 30.69 & 0.453 \\
    SpeechTokenizer & 40.62/70.08        & 80.82         & 57.70 & 51.97 & 41.82          & 60.33 & 0.143 & 51.03 & 0.476 \\
    WavLM           & 4.02/6.55          & 6.00          & \textbf{64.06} & \textbf{64.62} & 68.67 & \textbf{69.89} & 0.164 & \textbf{61.72} & \textbf{0.545} \\
    DAC             & 171.02/177.48      & 208.46        & 47.46 & 36.24 & 7.90           & 51.14 & \textbf{0.211} & 35.17 & 0.478 \\
    \midrule
    
    \multicolumn{10}{c}{\textit{8B Models}} \\
    \midrule
    HuBERT     & \textbf{2.24/4.38} & \textbf{3.13} & 47.62 & 52.21 & \textbf{70.13} & 48.03 & 0.129 & 45.19 & 0.507 \\
    SpeechTokenizer & 18.33/38.75        & 83.40         & 56.49 & 54.30 & 41.47          & \textbf{71.92} & 0.083 & 51.72 & 0.392 \\
    WavLM           & 2.83/5.57          & 5.49          & \textbf{62.29} & \textbf{67.81} & 65.39 & 71.45 & \textbf{0.177} & \textbf{59.31} & \textbf{0.523} \\
    DAC             & 147.35/148.41      & 166.57        & 46.01 & 29.85 & 7.42           & 48.86 & 0.101 & 36.55 & 0.501 \\
    \bottomrule
  \end{tabular}
  }
\end{table*}

\subsection{Scaling, Data Diversity, and Efficiency}

\textbf{Backbone Scaling and the Inverse Scaling Phenomenon.} We analyze scaling behavior separately within each model family: SmolLM2 (135M to 360M) and Llama-3 (1B to 8B). As shown in Tables~\ref{tab:encoder_decoder_vertical} and \ref{tab:main_results}, increasing capacity does not yield universal improvements. For SmolLM2, scaling from 135M to 360M provides marginal gains or fluctuations, suggesting capacity saturation for the evaluated encoders. For Llama-3, while scaling laws~\cite{kaplan2020scaling} hold for the data-abundant ASR task (LS-960), performance regression occurs in many data-limited scenarios. Overall, across both frameworks, performance decreased in approximately 53\% of the evaluated encoder-task configurations. These results indicate that for audio understanding, scaling backbones cannot compensate for low feature quality or limited training data.
While we focus on representations rather than projector design, our simple MLP might act as a bottleneck limiting these scaling benefits, which future context-aware architectures could mitigate by preventing surjective mappings.

\textbf{Impact of Pre-training Data Diversity.} Encoder understanding performance depends on the breadth of pre-training.
While semantic compatibility governs overall efficacy, the task-level breakdowns in Tables \ref{tab:encoder_decoder_vertical} and \ref{tab:main_results} further elucidate the critical impact of pre-training data diversity.
As observed in Table \ref{tab:encoder_decoder_vertical}, encoders pre-trained on diverse, multi-domain data (Whisper and WavLM) show a clear advantage over those trained exclusively on speech (HuBERT and SpeechTokenizer). Furthermore, results in Table \ref{tab:main_results} provide a more nuanced perspective: although HuBERT excels in tasks heavily reliant on speech-specific semantics, such as ASR and IC—sometimes even surpassing WavLM—the latter maintains its dominance across most of other general-audio categories. This gap underscores that multi-domain pre-training is essential for universal audio understanding, since instruction-tuning alone cannot bridge representation gaps inherited from narrow pre-training data.

\begin{figure}[th]
  \centering
  \includegraphics[width=0.9\linewidth]{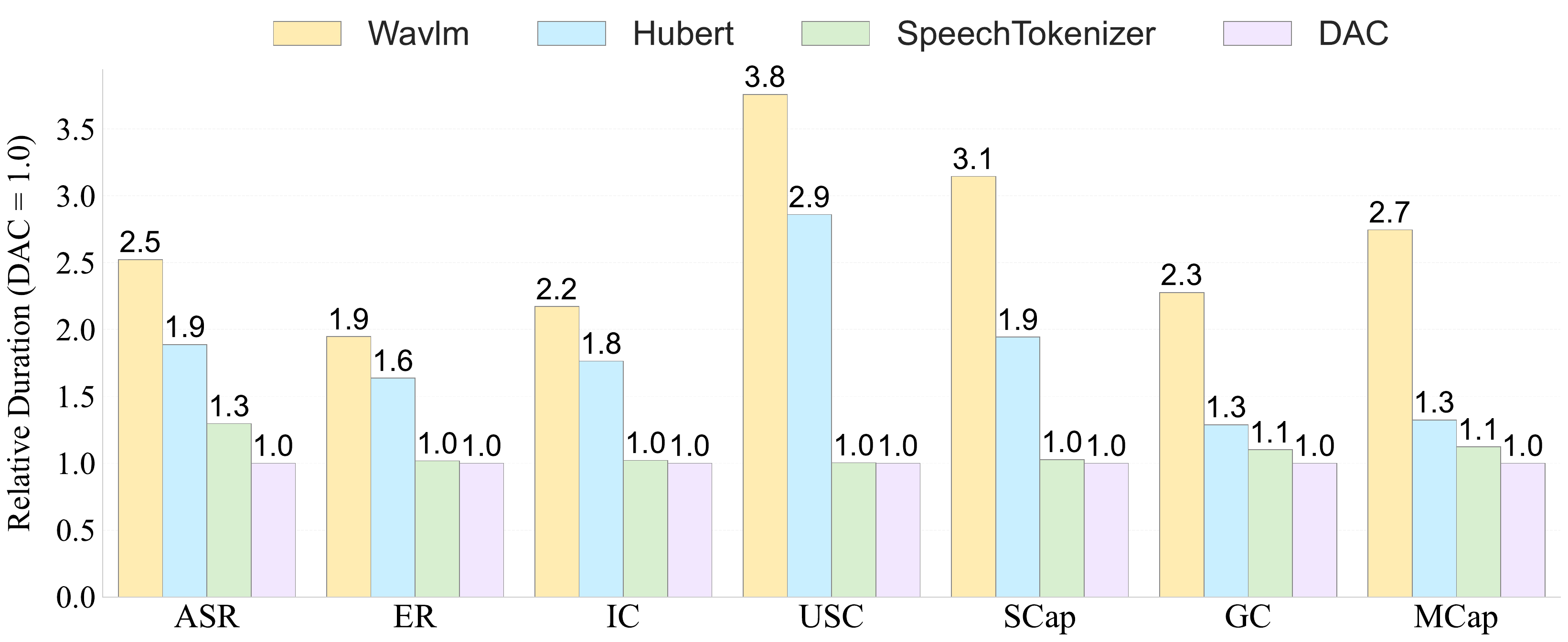}
  \caption{Total training time until convergence for encoders.}
  \label{fig:time}
\end{figure}

\textbf{Computational Efficiency and Training Convergence.} 
As illustrated in Figure~\ref{fig:time}, the computational cost of continuous features is a practical challenge. The sequence processing latency for continuous encoders (HuBERT and WavLM) significantly exceeds that of discrete tokenizers (SpeechTokenizer and DAC) across all benchmarks. Notably, discrete representations converge much faster during modality alignment, with validation performance stabilizing in fewer training steps. Despite the performance edge of continuous features in certain domains, the rapid convergence and lower overhead of discrete tokens make them an attractive paradigm for resource-constrained research and rapid system deployment.

\section{Conclusion}

This paper systematically evaluates continuous and discrete representations for audio understanding across speech, sound, and music. Results reveal that discrete and continuous encoders exhibit distinct and complementary strengths across task requirements. We highlight the necessity of incorporating semantic constraints into discrete representations.
Furthermore, our study underscores multi-domain pre-training data in achieving universal audio intelligence. Crucially, for audio understanding, backbone scaling fails to consistently compensate for an encoder's low semantic density, often leading to performance regression in data-limited tasks. These insights guide future LALM development, suggesting a focus on balancing semantics and fidelity through multi-domain pre-training.

\section{Acknowledgments}
This work is supported by National Natural Science Foundation of China (62076144) and Shenzhen Science and Technology Program (JCYJ20220818101014030).

\section{Generative AI Use Disclosure}
During the preparation of this manuscript, the authors used generative AI tools exclusively for the purpose of language editing and manuscript polishing to improve readability. These tools were not used to generate any core scientific ideas, experimental data, or technical contributions. All authors have thoroughly reviewed and approved the final version of the manuscript, and assume full responsibility for the integrity and entirety of its content.

\bibliographystyle{IEEEtran}
\bibliography{mybib}

@article{hsu2021hubert,
  title={Hubert: Self-supervised speech representation learning by masked prediction of hidden units},
  author={Hsu, Wei-Ning and Bolte, Benjamin and Tsai, Yao-Hung Hubert and Lakhotia, Kushal and Salakhutdinov, Ruslan and Mohamed, Abdelrahman},
  journal={IEEE/ACM Transactions on Audio, Speech, and Language Processing (TASLP)},
  pages={3451--3460},
  year={2021},
  publisher={IEEE}
}

@article{chen2022wavlm,
  title={WavLM: Large-Scale Self-Supervised Pre-Training for Full Stack Speech Processing},
  author={Chen, Sanyuan and Wang, Chengyi and Chen, Zhengyang and Wu, Yu and Liu, Shujie and Chen, Zhuo and Li, Jinyu and Kanda, Naoyuki and Yoshioka, Takuya and Xiao, Xiong and others},
  journal={IEEE Journal of Selected Topics in Signal Processing (JSTSP)},
  pages={1505},
  year={2022}
}

@inproceedings{zhang2023speechtokenizer,
title={SpeechTokenizer: Unified Speech Tokenizer for Speech Language Models},
author={Xin Zhang and Dong Zhang and Shimin Li and Yaqian Zhou and Xipeng Qiu},
booktitle={International Conference on Learning Representations (ICLR)},
year={2024},
}

@inproceedings{DAC,
  title={High-Fidelity Audio Compression with Improved RVQGAN},
  author={Kumar, Rithesh and Seetharaman, Prem and Luebs, Alejandro and Kumar, Ishaan and Kumar, Kundan},
  journal={Advances in Neural Information Processing Systems (NeurIPS)},
  pages={27980--27993},
  year={2023}
}

@inproceedings{wang2025speech,
  title={Speech Discrete Tokens or Continuous Features? A Comparative Analysis for Spoken Language Understanding in SpeechLLMs},
  author={Wang, Dingdong and Li, Junan and Cui, Mingyu and Yang, Dongchao and Chen, Xueyuan and Meng, Helen},
  booktitle={Empirical Methods in Natural Language Processing (EMNLP)},
  pages={24924--24935},
  year={2025}
}

@inproceedings{xu2024comparing,
  title={Comparing Discrete and Continuous Space LLMs for Speech Recognition},
  author={Xu, Yaoxun and Zhang, Shi-Xiong and Yu, Jianwei and Wu, Zhiyong and Yu, Dong},
  booktitle={Annual Conference of the International Speech Communication Association (INTERSPEECH)},
  pages={2509--2513},
  year={2024},
  organization={ISCA}
}

@article{mousavi2024dasb,
  title={DASB - Discrete Audio and Speech Benchmark},
  author={Mousavi, Pooneh and Della Libera, Luca and Duret, Jarod and Ploujnikov, Artem and Subakan, Cem and Ravanelli, Mirco},
  journal={arXiv preprint arXiv:2406.14294},
  year={2024}
}

@inproceedings{zhang2025x,
  title={X-ARES: A Comprehensive Framework for Assessing Audio Encoder Performance},
  author={Zhang, Junbo and Dinkel, Heinrich and Niu, Yadong and Liu, Chenyu and Cheng, Si and Zhao, Anbei and Luan, Jian},
  booktitle={Annual Conference of the International Speech Communication Association (INTERSPEECH)},
  pages={4868--4872},
  organization={ISCA},
  year={2025}
}

@article{mousavi2025discrete,
  title={Discrete Audio Tokens: More Than a Survey!},
  author={Mousavi, Pooneh and Maimon, Gallil and Moumen, Adel and Petermann, Darius and Shi, Jiatong and Wu, Haibin and Yang, Haici and Kuznetsova, Anastasia and Ploujnikov, Artem and Marxer, Ricard and others},
  journal={Transactions on Machine Learning Research (TMLR)},
  year={2025},
  publisher={Transactions on Machine Learning Research}
}

@article{guo2025recent,
  title={Recent Advances in Discrete Speech Tokens: A Review},
  author={Guo, Yiwei and Li, Zhihan and Wang, Hankun and Li, Bohan and Shao, Chongtian and Zhang, Hanglei and Du, Chenpeng and Chen, Xie and Liu, Shujie and Yu, Kai},
  journal={IEEE Transactions on Pattern Analysis and Machine Intelligence (TPAMI)}
}

@article{zhou2025voxcpm,
  title={VoxCPM: Tokenizer-Free TTS for Context-Aware Speech Generation and True-to-Life Voice Cloning},
  author={Zhou, Yixuan and Zeng, Guoyang and Liu, Xin and Li, Xiang and Yu, Renjie and Wang, Ziyang and Ye, Runchuan and Sun, Weiyue and Gui, Jiancheng and Li, Kehan and others},
  journal={arXiv preprint arXiv:2509.24650},
  year={2025}
}

@inproceedings{allal2025smollm,
    title={Smol{LM}2: When Smol Goes Big {\textemdash} Data-Centric Training of a Fully Open Small Language Model},
    author={Loubna Ben allal and Anton Lozhkov and Elie Bakouch and others},
    booktitle={Second Conference on Language Modeling (COLM)},
    year={2025}
}

@article{grattafiori2024llama,
  title={The Llama 3 Herd of Models},
  author={Grattafiori, Aaron and Dubey, Abhimanyu and Jauhri, Abhinav and Pandey, Abhinav and Kadian, Abhishek and Al-Dahle, Ahmad and Letman, Aiesha and Mathur, Akhil and Schelten, Alan and Vaughan, Alex and others},
  journal={arXiv preprint arXiv:2407.21783},
  year={2024}
}

@inproceedings{panayotov2015librispeech,
  title={Librispeech: An ASR Corpus Based on Public Domain Audio Books},
  author={Panayotov, Vassil and Chen, Guoguo and Povey, Daniel and Khudanpur, Sanjeev},
  booktitle={IEEE International Conference on Acoustics, Speech and Signal Processing (ICASSP)},
  pages={5206--5210},
  year={2015},
  organization={IEEE}
}

@article{busso2008iemocap,
  title={{IEMOCAP}: Interactive Emotional Dyadic Motion Capture Database},
  author={Busso, Carlos and Bulut, Murtaza and Lee, Chi-Chun and Kazemzadeh, Abe and Mower, Emily and Kim, Samuel and Chang, Jeannette N and Lee, Sungbok and Narayanan, Shrikanth S},
  journal={Language Resources and Evaluation (LRE)},
  pages={335--359},
  year={2008},
  publisher={Springer}
}

@inproceedings{bastianelli2020slurp,
  title={SLURP: A Spoken Language Understanding Resource Package},
  author={Bastianelli, Emanuele and Vanzo, Andrea and Swietojanski, Pawel and Rieser, Verena},
  booktitle={Empirical Methods in Natural Language Processing (EMNLP)},
  pages={7252--7262},
  year={2020}
}

@inproceedings{manco2023song,
  title={The Song Describer Dataset: A Corpus of Audio Captions for Music-and-Language Evaluation},
  author={Manco, Ilaria and Weck, Benno and Doh, Seungheon and Won, Minz and Zhang, Yixiao and Bogdanov, Dmitry and Wu, Yusong and Chen, Ke and Tovstogan, Philip and Benetos, Emmanouil and others},
  booktitle={Workshop on Machine Learning for Audio, Neural Information Processing Systems (NeurIPS)},
  year={2023},
  organization={Neural Information Processing Systems}
}

@inproceedings{baevski2020wav2vec,
  title={Wav2vec 2.0: A Framework for Self-Supervised Learning of Speech Representations},
  author={Baevski, Alexei and Zhou, Yuhao and Mohamed, Abdelrahman and Auli, Michael},
  booktitle={Advances in Neural Information Processing Systems (NeurIPS)},
  pages={12449--12460},
  year={2020}
}

@inproceedings{radford2023robust,
  title={Robust Speech Recognition via Large-Scale Weak Supervision},
  author={Radford, Alec and Kim, Jong Wook and Xu, Tao and Brockman, Greg and McLeavey, Christine and Sutskever, Ilya},
  booktitle={International Conference on Machine Learning (ICML)},
  pages={28492--28518},
  year={2023},
  organization={PMLR}
}

@inproceedings{ji2024wavtokenizer,
  title={WavTokenizer: An Efficient Acoustic Discrete Codec Tokenizer for Audio Language Modeling},
  author={Ji, Shengpeng and Jiang, Ziyue and Wang, Wen and Chen, Yifu and Fang, Minghui and Zuo, Jialong and Yang, Qian and Cheng, Xize and Wang, Zehan and Li, Ruiqi and others},
  booktitle={International Conference on Learning Representations (ICLR)}
}

@inproceedings{salamon2014dataset,
  title={A Dataset and Taxonomy for Urban Sound Research},
  author={Salamon, Justin and Jacoby, Christopher and Bello, Juan Pablo},
  booktitle={ACM international conference on Multimedia (MM)},
  pages={1041--1044},
  year={2014}
}

@inproceedings{tzanetakis2001automatic,
  title={Automatic Musical Genre Classification Of Audio Signals},
  author={Tzanetakis, George and Essl, Georg and Cook, Perry},
  booktitle={International Symposium on Music Information Retrieval (ISMIR)},
  year={2001}
}

@inproceedings{engel2017neural,
  title={Neural Audio Synthesis of Musical Notes with WaveNet Autoencoders},
  author={Engel, Jesse and Resnick, Cinjon and Roberts, Adam and Dieleman, Sander and Eck, Douglas and Simonyan, Karen and Norouzi, Mohammad},
  booktitle={International Conference on Machine Learning (ICML)},
  pages={1068--1077},
  year={2017},
  organization={JMLR. org}
}

@inproceedings{drossos2020clotho,
  title={Clotho: An Audio Captioning Dataset},
  author={Drossos, Konstantinos and Lipping, Samuel and Virtanen, Tuomas},
  booktitle={IEEE International Conference on Acoustics, Speech and Signal Processing (ICASSP)},
  pages={736--740},
  year={2020},
  organization={IEEE}
}

@article{cao2014crema,
  title={{CREMA-D}: Crowd-Sourced Emotional Multimodal Actors Dataset},
  author={Cao, Houwei and Cooper, David G and Keutmann, Michael K and Gur, Ruben C and Nenkova, Ani and Verma, Ragini},
  journal={IEEE Transactions on Affective Computing (TAC)},
  pages={377--390},
  year={2014},
  publisher={IEEE}
}

@article{chu2023qwen,
  title={Qwen-Audio: Advancing Universal Audio Understanding via Unified Large-Scale Audio-Language Models},
  author={Chu, Yunfei and Xu, Jin and Zhou, Xiaohuan and Yang, Qian and Zhang, Shiliang and Yan, Zhijie and Zhou, Chang and Zhou, Jingren},
  journal={arXiv preprint arXiv:2311.07919},
  year={2023}
}

@inproceedings{tang2023salmonn,
  title={SALMONN: Towards Generic Hearing Abilities for Large Language Models},
  author={Tang, Changli and Yu, Wenyi and Sun, Guangzhi and Chen, Xianzhao and Tan, Tian and Li, Wei and Lu, Lu and MA, Zejun and Zhang, Chao},
  booktitle={International Conference on Learning Representations (ICLR)}
}

@article{hu2022lora,
  title={LoRA: Low-Rank Adaptation of Large Language Models},
  author={Hu, Edward J and Shen, Yelong and Wallis, Phillip and Allen-Zhu, Zeyuan and Li, Yuanzhi and Wang, Shean and Wang, Liang and Chen, Weizhu and others},
  journal={International Conference on Learning Representations (ICLR)},
  pages={3},
  year={2022}
}

@article{borsos2023audiolm,
  title={AudioLM: A Language Modeling Approach to Audio Generation},
  author={Borsos, Zal{\'a}n and Marinier, Rapha{\"e}l and Vincent, Damien and Kharitonov, Eugene and Pietquin, Olivier and Sharifi, Matt and Roblek, Dominik and Teboul, Olivier and Grangier, David and Tagliasacchi, Marco and others},
  journal={IEEE/ACM Transactions on Audio, Speech, and Language Processing (TASLP)},
  pages={2523--2533},
  year={2023},
  publisher={IEEE}
}

@inproceedings{zhang2023speechgpt,
  title={SpeechGPT: Empowering Large Language Models with Intrinsic Cross-Modal Conversational Abilities},
  author={Zhang, Dong and Li, Shimin and Zhang, Xin and Zhan, Jun and Wang, Pengyu and Zhou, Yaqian and Qiu, Xipeng},
  booktitle={Findings of the Association for Computational Linguistics: EMNLP 2023 (EMNLP Findings)},
  pages={15757--15773},
  year={2023}
}

@INPROCEEDINGS{zhou2022can,
  author={Zhou, Zelin and Zhang, Zhiling and Xu, Xuenan and Xie, Zeyu and Wu, Mengyue and Zhu, Kenny Q.},
  booktitle={IEEE International Conference on Acoustics, Speech and Signal Processing (ICASSP)}, 
  title={Can Audio Captions Be Evaluated With Image Caption Metrics?}, 
  year={2022},
  pages={981--985},
}

@inproceedings{ravanelli2019speech,
    title={Speech Model Pre-Training for End-to-End Spoken Language Understanding},
    author={Ravanelli, Mirco and Tomar, Vikrant and Ignoto, Patrick and Rachmad, Yoesoep and Lugosch, Loren},
    booktitle={Annual Conference of the International Speech Communication Association (INTERSPEECH)},
    pages={814--818},
    organization={ISCA},
    year={2019},
}

@article{warden2018speech,
  title={Speech Commands: A Dataset for Limited-Vocabulary Speech Recognition},
  author={Warden, Pete},
  journal={arXiv preprint arXiv:1804.03209},
  year={2018},
}

@article{nagrani2019voxceleb,
  title={VoxCeleb: Large-scale Speaker Verification in the Wild},
  author={Nagrani, Arsha and Chung, Joon Son and Xie, Weidi and Zisserman, Andrew},
  journal={Computer Speech \& Language (CSL)},
  volume={60},
  pages={101027},
  year={2019},
}

@article{Valk2020VOXLINGUA107AD,
  title={VOXLINGUA107: A Dataset for Spoken Language Recognition},
  author={J{\"o}rgen Valk and Tanel Alum{\"a}e},
  journal={2021 IEEE Spoken Language Technology Workshop (SLT)},
  year={2020},
  pages={652-658},
}

@article{stoter2018libricount,
  title={LibriCount, a dataset for speaker count estimation},
  author={St{\"{o}}ter, Fabian-Robert and Chakrabarty, Soumitro and Habets, Emanu{\"e}l and Edler, Bernd},
  year={2018},
  publisher={Zenodo}
}

@inproceedings{gong2022vocalsound,
  title={Vocalsound: A dataset for improving human vocal sounds recognition},
  author={Gong, Yuan and Yu, Jin and Glass, James},
  booktitle={IEEE International Conference on Acoustics, Speech and Signal Processing (ICASSP)},
  pages={151--155},
  year={2022},
}

@article{wu2017asvspoof,
  title={ASVspoof: The automatic speaker verification spoofing and countermeasures challenge},
  author={Wu, Zhizheng and Yamagishi, Junichi and Kinnunen, Tomi and Hanil{\c{c}}i, Cemal and Sahidullah, Mohammed and Sizov, Aleksandr and Evans, Nicholas and Todisco, Massimiliano and Delgado, Hector},
  journal={IEEE Journal of Selected Topics in Signal Processing (JSTSP)},
  volume={11},
  number={4},
  pages={588--604},
  year={2017},
}

@inproceedings{piczak2015esc,
  title={ESC: Dataset for environmental sound classification},
  author={Piczak, Karol J},
  booktitle={ACM international conference on Multimedia (MM)},
  pages={1015--1018},
  year={2015}
}

@article{fonseca2021fsd50k,
  title={{FSD50K}: an open dataset of human-labeled sound events},
  author={Fonseca, Eduardo and Favory, Xavier and Pons, Jordi and Font, Frederic and Serra, Xavier},
  journal={IEEE/ACM Transactions on Audio, Speech, and Language Processing (TASLP)},
  volume={30},
  pages={829--852},
  year={2021},
}

@inproceedings{fonseca2018general,
  title={General-Purpose Tagging of Freesound Audio with AudioSet Labels: Task Description, Dataset and Baseline},
  author={Fonseca, Eduardo and Plakal, Manoj and Font, Frederic and Ellis, Daniel P. W. and Favory, Xavier and Pons, Jordi and Serra, Xavier},
  booktitle={Detection and Classification of Acoustic Scenes and Events Workshop (DCASE)},
  year={2018}
}

@inproceedings{Defferrard2016FMAAD,
  title={FMA: A Dataset for Music Analysis},
  author={Micha{\"e}l Defferrard and Kirell Benzi and Pierre Vandergheynst and Xavier Bresson},
  booktitle={International Society for Music Information Retrieval Conference (ISMIR)},
  year={2016},
}

@article{niu2025mecat,
  title={MECAT: A Multi-Experts Constructed Benchmark for Fine-Grained Audio Understanding Tasks},
  author={Niu, Yadong and Wang, Tianzi and Dinkel, Heinrich and Sun, Xingwei and Zhou, Jiahao and Li, Gang and Liu, Jizhong and Liu, Xunying and Zhang, Junbo and Luan, Jian},
  journal={arXiv preprint arXiv:2507.23511},
  year={2025}
}

@inproceedings{bu2017aishell,
  title={Aishell-1: An open-source mandarin speech corpus and a speech recognition baseline},
  author={Bu, Hui and Du, Jiayu and Na, Xingyu and Wu, Bengu and Zheng, Hao},
  booktitle={conference of the oriental chapter of the international coordinating committee on speech databases and speech I/O systems and assessment (O-COCOSDA)},
  pages={1--5},
  year={2017},
  organization={IEEE}
}

@inproceedings{
    loshchilov2018decoupled,
    title={Decoupled Weight Decay Regularization},
    author={Ilya Loshchilov and Frank Hutter},
    booktitle={International Conference on Learning Representations (ICLR)},
    year={2019},
}

@article{kaplan2020scaling,
  title={Scaling Laws for Neural Language Models},
  author={Kaplan, Jared and McCandlish, Sam and Henighan, Thomas and Brown, Tom B. and Chess, Benjamin and Child, Rewon and Gray, Scott and Radford, Alec and Wu, Jeff and Amodei, Dario},
  journal={arXiv preprint arXiv:2001.08361},
  year={2020},
}

\end{document}